\documentstyle[12pt,epsf]{article}
\newlength{\pubnumber} 

\catcode`\@=11
\@addtoreset{equation}{section}

\def\section{\@startsection{section}{1}{\z@}{3.5ex plus 1ex minus .2ex}
 {2.3ex plus .2ex}{\large\bf}}
\def\subsection{\@startsection{subsection}{2}{\z@}{2.3ex plus .2ex}
 {2.3ex plus .2ex}{\bf}}

\input epsf
\newcommand{\oo}[2]{\left(#1\left|#2\right.\right)}
\newcommand{\ba}{\begin{eqnarray}}
\newcommand{\ea}{\end{eqnarray}}
\begin{document}

\begin{titlepage}
\samepage{
\setcounter{page}{1}
\rightline{LPTENS--06/55}
\rightline{LTH--715}

\rightline{\tt hep-th/0611251}
\rightline{June 2006}
\vfill
\begin{center}
 {\Large \bf
Spinor--Vector Duality \\ in \\
fermionic $Z_2\times Z_2$ heterotic orbifold models
}
\vspace{1cm}
\vfill {\large Alon E. Faraggi$^{1}$,
Costas Kounnas$^{2}$\footnote{Unit\'e Mixte de Recherche
(UMR 8549) du CNRS et de l'ENS
 associ\'e©e a l'universit\'e© Pierre et Marie Curie}
 and
John Rizos$^{3}$}\\
\vspace{1cm}
{\it $^{1}$ Dept.\ of Mathematical Sciences,
             University of Liverpool,
         Liverpool L69 7ZL, UK\\}
\vspace{.05in}
{\it $^{2}$ Lab.\ Physique Th\'eorique,
Ecole Normale Sup\'erieure, F--75231 Paris 05, France\\}
\vspace{.05in}
{\it $^{3}$ Department of Physics,
              University of Ioannina, GR45110 Ioannina, Greece\\}
\vspace{.025in}
\end{center}
\vfill
\begin{abstract}
We continue the classification of the fermionic $Z_2\times Z_2$
heterotic string vacua with symmetric internal shifts.
The space of models is spanned by working with a fixed set of
boundary condition basis vectors and by varying the sets of
independent Generalized GSO (GGSO) projection coefficients
(discrete torsion). This includes the Calabi--Yau like compactifications
with (2,2) world--sheet superconformal symmetry, as well as
more general vacua with only (2,0) superconformal symmetry.
In contrast to our earlier classification that utilized
a montecarlo technique to generate random sets of GGSO
phases, in this paper we present the results of a complete
classification of the subclass of the models in which the
four dimensional gauge group arises solely from the null sector.
In line with the results of the statistical classification
we find a bell shaped distribution that peaks at vanishing
net number of generations and with $\sim$15\% of the models
having three net chiral families. The complete classification reveals
a novel spinor--vector
duality symmetry over the entire space of vacua. The $S_t\leftrightarrow V$
duality interchanges the spinor plus anti--spinor representations
with vector representations. We present the data that
demonstrates the spinor--vector duality.
We illustrate the existence of a duality map in
a concrete example. We provide a general algebraic proof for the
existence of the $S_t\leftrightarrow V$ duality map.
We discuss the case of self--dual solutions with an equal
number of vectors and spinors, in the presence and absence of
$E_6$ gauge symmetry,
and presents a couple of concrete examples of self--dual models
without $E_6$ symmetry.

\noindent

\end{abstract}
\smallskip}
\end{titlepage}

\setcounter{footnote}{0}

\def\beq{\begin{equation}}
\def\eeq{\end{equation}}
\def\beqn{\begin{eqnarray}}
\def\eeqn{\end{eqnarray}}

\def\no{\noindent }
\def\nolabel{\nonumber }
\def\ie{{\it i.e.}}
\def\eg{{\it e.g.}}
\def\half{{\textstyle{1\over 2}}}
\def\third{{\textstyle {1\over3}}}
\def\quarter{{\textstyle {1\over4}}}
\def\sixth{{\textstyle {1\over6}}}
\def\m{{\tt -}}
\def\p{{\tt +}}

\def\Tr{{\rm Tr}\, }
\def\tr{{\rm tr}\, }

\def\slash#1{#1\hskip-6pt/\hskip6pt}
\def\slk{\slash{k}}
\def\GeV{\,{\rm GeV}}
\def\TeV{\,{\rm TeV}}
\def\y{\,{\rm y}}
\def\SM{Standard--Model }
\def\SUSY{supersymmetry }
\def\SSSM{supersymmetric standard model}
\def\vev#1{\left\langle #1\right\rangle}
\def\l{\langle}
\def\r{\rangle}
\def\o#1{\frac{1}{#1}}

\def\Htw{{\tilde H}}
\def\chibar{{\overline{\chi}}}
\def\qbar{{\overline{q}}}
\def\ibar{{\overline{\imath}}}
\def\jbar{{\overline{\jmath}}}
\def\Hbar{{\overline{H}}}
\def\Qbar{{\overline{Q}}}
\def\abar{{\overline{a}}}
\def\alphabar{{\overline{\alpha}}}
\def\betabar{{\overline{\beta}}}
\def\tautwo{{ \tau_2 }}
\def\thetatwo{{ \vartheta_2 }}
\def\thetathree{{ \vartheta_3 }}
\def\thetafour{{ \vartheta_4 }}
\def\ttwo{{\vartheta_2}}
\def\tthree{{\vartheta_3}}
\def\tfour{{\vartheta_4}}
\def\ti{{\vartheta_i}}
\def\tj{{\vartheta_j}}
\def\tk{{\vartheta_k}}
\def\calF{{\cal F}}
\def\smallmatrix#1#2#3#4{{ {{#1}~{#2}\choose{#3}~{#4}} }}
\def\ab{{\alpha\beta}}
\def\Minv{{ (M^{-1}_\ab)_{ij} }}
\def\bone{{\bf 1}}
\def\ii{{(i)}}
\def\V{{\bf V}}
\def\N{{\bf N}}

\def\b{{\bf b}}
\def\S{{\bf S}}
\def\X{{\bf X}}
\def\I{{\bf I}}
\def\mb{{\mathbf b}}
\def\mS{{\mathbf S}}
\def\mX{{\mathbf X}}
\def\mI{{\mathbf I}}
\def\balpha{{\mathbf \alpha}}
\def\bbeta{{\mathbf \beta}}
\def\bgamma{{\mathbf \gamma}}
\def\bxi{{\mathbf \xi}}

\def\t#1#2{{ \Theta\left\lbrack \matrix{ {#1}\cr {#2}\cr }\right\rbrack }}
\def\C#1#2{{ C\left\lbrack \matrix{ {#1}\cr {#2}\cr }\right\rbrack }}
\def\tp#1#2{{ \Theta'\left\lbrack \matrix{ {#1}\cr {#2}\cr }\right\rbrack }}
\def\tpp#1#2{{ \Theta''\left\lbrack \matrix{ {#1}\cr {#2}\cr }\right\rbrack }}
\def\l{\langle}
\def\r{\rangle}
\newcommand{\cc}[2]{c{#1\atopwithdelims[]#2}}
\newcommand{\nn}{\nonumber}


\def\inbar{\,\vrule height1.5ex width.4pt depth0pt}

\def\IC{\relax\hbox{$\inbar\kern-.3em{\rm C}$}}
\def\IQ{\relax\hbox{$\inbar\kern-.3em{\rm Q}$}}
\def\IR{\relax{\rm I\kern-.18em R}}
 \font\cmss=cmss10 \font\cmsss=cmss10 at 7pt
\def\IZ{\relax\ifmmode\mathchoice
 {\hbox{\cmss Z\kern-.4em Z}}{\hbox{\cmss Z\kern-.4em Z}}
 {\lower.9pt\hbox{\cmsss Z\kern-.4em Z}}
 {\lower1.2pt\hbox{\cmsss Z\kern-.4em Z}}\else{\cmss Z\kern-.4em Z}\fi}

\def\AEF{A.E. Faraggi}
\def\JHEP#1#2#3{{\it JHEP}\/ {\bf #1} (#2) #3}
\def\NPB#1#2#3{{\it Nucl.\ Phys.}\/ {\bf B#1} (#2) #3}
\def\PLB#1#2#3{{\it Phys.\ Lett.}\/ {\bf B#1} (#2) #3}
\def\PRD#1#2#3{{\it Phys.\ Rev.}\/ {\bf D#1} (#2) #3}
\def\PRL#1#2#3{{\it Phys.\ Rev.\ Lett.}\/ {\bf #1} (#2) #3}
\def\PRT#1#2#3{{\it Phys.\ Rep.}\/ {\bf#1} (#2) #3}
\def\MODA#1#2#3{{\it Mod.\ Phys.\ Lett.}\/ {\bf A#1} (#2) #3}
\def\IJMP#1#2#3{{\it Int.\ J.\ Mod.\ Phys.}\/ {\bf A#1} (#2) #3}
\def\nuvc#1#2#3{{\it Nuovo Cimento}\/ {\bf #1A} (#2) #3}
\def\RPP#1#2#3{{\it Rept.\ Prog.\ Phys.}\/ {\bf #1} (#2) #3}
\def\etal{{\it et al\/}}

\hyphenation{su-per-sym-met-ric non-su-per-sym-met-ric}
\hyphenation{space-time-super-sym-met-ric}
\hyphenation{mod-u-lar mod-u-lar--in-var-i-ant}


\setcounter{footnote}{0}
\section{Introduction}

In the framework of the free fermionic construction \cite{fff1,fff2}
of the heterotic string many three generation realistic string models can be
constructed in four dimensions with the correct quantum numbers under
the Standard Model gauge group \cite{ffm}. In the orbifold language
the free fermionic models corresponds to a
either symmetric or asymmetric, or freely acting,  orbifolds.
In particular, a subclass of the free fermionic vacua correspond to
symmetric $Z_2\times Z_2$ orbifold compactifications at enhanced symmetry
points in the toroidal moduli space \cite{z2z21,z2z22}.
In this class of orbifold models
the chiral matter spectrum arises from twisted sectors and
thus does not depend on the moduli.
This allows the development of a complete classification of
$Z_2\times Z_2$ symmetric orbifolds. The free fermionic construction provides
the techniques which facilitate developing a computerized classification
algorithm for the twisted matter chiral spectrum.
Thus, the free fermionic formalism provides powerful tools for the
systematic classification of symmetric $Z_2\times Z_2$ perturbative
string orbifold models.

For type II string $N=2$ supersymmetric vacua
the general free fermionic classification techniques were developed in ref.
\cite{gkr}. The method was extended in refs.
\cite{fknr,nooij,fkr} for the classification of heterotic $Z_2\times Z_2$
orbifolds.
In this class of models the six dimensional internal manifold
contains three twisted sectors.
In the heterotic string each of these sectors
may, or may not, a priori
(prior to application of the Generalized GSO (GGSO) projections),
give rise to spinorial representations.
Generically we may classify the models as
$S^3$, $S^2V$, $SV^2$ and $V^3$ classes of models, with spinorial
representations arising from three, two, one or none of the twisted sectors,
respectively.
A priori it may be thought that the classification of the
different classes of models would require different sets of basis vectors.
However, In ref. \cite{fkr} we demonstrated that
the entire sets of $S^3,~S^2V,~SV^2$ and $V^3$ classes of models
are produced by
working with the single basis set of ref. \cite{fknr},
according to specific choices of the one--loop Generalized GSO
(GGSO) projection coefficients (discrete torsions).
This fact is of basic importance since it
enables a systematic analysis of {\it all} the models and
the representation of their main features,
like the number of spinorial, anti--spinorial and vectorial
representations, in algebraic formulas.

The classification methodology that
we developed allows us to scan a range of over
$10^{16}$ models,
and therefore obtain vital insight into the properties
of the entire space of symmetric $Z_2\times Z_2$ orbifold vacua.
The space of vacua in this class of models arises from a set of
independent binary GGSO projection coefficients  $\cc{b_i}{b_j}$, which
correspond a matrix with elements taking values $\pm1$.
The independent elements of this matric correspond to the
upper block of this matrix. All other elements are fixed
by modular invariance and factorization of the partition function
\cite{fff1}. The classification basis of ref. \cite{fknr}
contains 12 vectors. Therefore, the number of independent GGSO
projection coefficients is 66.
Requiring $N=1$ space--time supersymmetry reduces the number of independent
phases to 55. Hence, prior to imposing further constraints
the space of models that we scan contains $2^{55}$ different vacua.
This space of models is still too large for a complete
computerized classification.
Therefore, in ref. \cite{fkr} we resorted to a montecarlo technique
to generate random choices of phases. Our computer method
checked that only new models are recounted and in this manner
we were able to explore a set of some $10^{10}$ distinct vacua.

The analysis in ref. \cite{fkr} revealed a bell shape distribution
that peaks at vanishing net number of chiral families,
with about 15\% of the models having three net chiral families.
The statistical analysis also revealed an additional symmetry
in the distribution of $Z_2\times Z_2$ string vacua
under exchange of vectorial, and spinorial plus anti--spinorial,
representations of $SO(10)$. This symmetry is akin to mirror symmetry
which exchanges spinorial with anti--spinorial representations.
The symmetry is observed by noting that the same number of models
are generated under the exchange.

In this paper we continue the study of the classification with particular
focus on the exploration of the new symmetry under the exchange of spinorial
and vectorial representations. In particular, for this purpose we modify
the method
of analysis. Rather than using montecarlo generation of random sets of GGSO
phases sets, we perform a complete classification of a restricted class
of models. The restricted class is selected by imposing that the
only space-time vector bosons that arise in the models are those
that are obtained from the untwisted Neveu--Schwarz sector.
Vector bosons that may arise from other sectors are projected
out by the specific choice of GGSO projection coefficients.
This is achieved by restricting the choices of GGSO projection
coefficients and hence restricting the space of models, and enables
a complete computerized classification of the subclass of vacua.
This restricts the four dimensional gauge group in these models to be
$SO(10)\times U(1)^3 \times SO(8)\times SO(8)$, and eliminates
enhancements $SO(10)\times U(1)\rightarrow E_6$ as well as all enhancements of
the $SO(8)\times SO(8)$.

The complete classification of this restricted class
again reveals the symmetry under exchange
of the total number of spinors plus anti--spinors with the number of
vectors in the space of string vacua. Furthermore, we note that the symmetry
operates
on each of the three twisted sectors of the $Z_2\times Z_2$ orbifold.
We note that
the symmetry under this exchange is evident when the $SO(10)$ symmetry is
enhanced to $E_6$, in which case $\# (16+{\overline 16}) = \# (10)$.
We demonstrate that the symmetry persists also when
there is no enhancement to $E_6$.
We further show the existence of self--dual vacua in which
$\# (16+{\overline 16}) = \# (10)$, but in which the $SO(10)$ symmetry
is not enhanced to $E_6$.

Our paper is organised as follows: in section \ref{review} we review the method
of classification for completeness. In section \ref{count}
we elaborate on the counting method of $SO(10)$ spinorial and vectorial
representations. In section \ref{4dgroup} we discuss the conditions
imposed on the four dimensional gauge group and their implementation
in the classification method. In section \ref{results} we discuss
the results of the classification in comparison to the statistical
classification of ref. \cite{fkr}.
In section \ref{svduality} we discuss the spinor--vector duality.
In section \ref{proof} we provide an analytic proof of the spinor--vector
duality. In section \ref{sdsolutions} we demonstrate the existence of vacua
that are self--dual under the spinor--vector interchange, but in which the
$SO(10)$ symmetry is not enhanced to $E_6$. Section \ref{conclude}
concludes the paper.

\section{Review of the classification method}\label{review}

In the free fermionic formulation the 4-dimensional heterotic string,
in the light-cone gauge, is described
by $20$ left--moving  and $44$ right--moving two dimensional real
fermions \cite{fff1,fff2}.
A large number of models can be constructed by choosing
different phases picked up by   fermions ($f_A, A=1,\dots,44$) when transported
along
the torus non-contractible loops.
Each model corresponds to a particular choice of fermion phases consistent with
modular invariance
that can be generated by a set of  basis vectors $v_i,i=1,\dots,n$,
$$v_i=\left\{\alpha_i(f_1),\alpha_i(f_{2}),\alpha_i(f_{3}))\dots\right\}$$
describing the transformation  properties of each fermion
\begin{equation}
f_A\to -e^{i\pi\alpha_i(f_A)}\ f_A, \ , A=1,\dots,44~.
\end{equation}
The basis vectors span a space $\Xi$ which consists of $2^N$ sectors that give
rise to the string spectrum. Each sector is given by
\begin{equation}
\xi = \sum N_i v_i,\ \  N_i =0,1
\end{equation}
The spectrum is truncated by a GGSO projection whose action on a
string state  $|S>$ is
\begin{equation}\label{eq:gso}
e^{i\pi v_i\cdot F_S} |S> = \delta_{S}\ \cc{S}{v_i} |S>,
\end{equation}
where $F_S$ is the fermion number operator and $\delta_{S}=\pm1$ is the
spacetime spin statistics index.
Different sets of projection coefficients $\cc{S}{v_i}=\pm1$ consistent with
modular invariance give
rise to different models. Summarizing: a model can be defined uniquely by a set
of basis vectors $v_i,i=1,\dots,n$
and a set of $2^{N(N-1)/2}$ independent projections coefficients
$\cc{v_i}{v_j}, i>j$.

The two dimensional
free fermions in the light-cone gauge (in the usual notation
\cite{fff1,fff2,ffm}) are:
$\psi^\mu, \chi^i,y^i, \omega^i, i=1,\dots,6$ (real left-moving fermions)
and
$\bar{y}^i,\bar{\omega}^i, i=1,\dots,6$ (real right-moving fermions),
${\bar\psi}^A, A=1,\dots,5$, $\bar{\eta}^B, B=1,2,3$, $\bar{\phi}^\alpha,
\alpha=1,\ldots,8$ (complex right-moving fermions).
The class of models under investigation,
is generated by a set $V$ of 12 basis vectors
$$
V=\{v_1,v_2,\dots,v_{12}\},
$$
where
\begin{eqnarray}
v_1=1&=&\{\psi^\mu,\
\chi^{1,\dots,6},y^{1,\dots,6}, \omega^{1,\dots,6}| \nn\\
& & ~~~\bar{y}^{1,\dots,6},\bar{\omega}^{1,\dots,6},
\bar{\eta}^{1,2,3},
\bar{\psi}^{1,\dots,5},\bar{\phi}^{1,\dots,8}\},\nn\\
v_2=S&=&\{\psi^\mu,\chi^{1,\dots,6}\},\nn\\
v_{2+i}=e_i&=&\{y^{i},\omega^{i}|\bar{y}^i,\bar{\omega}^i\}, \
i=1,\dots,6,\nn\\
v_{9}=b_1&=&\{\chi^{34},\chi^{56},y^{34},y^{56}|\bar{y}^{34},
\bar{y}^{56},\bar{\eta}^1,\bar{\psi}^{1,\dots,5}\},\label{basis}\\
v_{10}=b_2&=&\{\chi^{12},\chi^{56},y^{12},y^{56}|\bar{y}^{12},
\bar{y}^{56},\bar{\eta}^2,\bar{\psi}^{1,\dots,5}\},\nn\\
v_{11}=z_1&=&\{\bar{\phi}^{1,\dots,4}\},\nn\\
v_{12}=z_2&=&\{\bar{\phi}^{5,\dots,8}\}.\nn
\end{eqnarray}
The vectors $1,S$ generate an
$N=4$ supersymmetric model, with $SO(44)$ gauge symmetry.
The vectors $e_i,i=1,\dots,6$ give rise
to all possible symmetric shifts of the six internal fermionized coordinates
($\partial X^i=y^i\omega^i, {\bar\partial} X^i= \bar{y}^i\bar{\omega}^i$).
Their addition breaks the $SO(44)$ gauge group, but preserves
$N=4$ supersymmetry.
The vectors $b_1$ and $b_2$
defines the $Z_2\times Z_2$ orbifold twists, which breaks
$N=4$ to $N=1$ supersymmetry, and defines the $SO(10)$ gauge symmetry.
The $z_1$ and $z_2$ basis vectors give rise to the $SO(8)\times SO(8)$
gauge group.
It is important to note here that the above
choice of $V$ is the most general set of
basis vectors, compatible with a $SO(10)$
Kac--Moody level one algebra.

Without loss of generality we can fix some of
the associated GGSO projection coefficients
$$
\cc{1}{1}=\cc{1}{S}=\cc{S}{S}=\cc{S}{e_i}=\cc{S}{b_A}=-
\cc{b_2}{S}=\cc{S}{z_n}=-1,~
$$
leaving 55 independent coefficients,
\begin{eqnarray}
&&\cc{e_i}{e_j}, i\ge j, \ \ \cc{b_1}{b_2}, \ \ \cc{z_1}{z_2},\nn\\
&&\cc{e_i}{z_n}, \cc{e_i}{b_A},\cc{b_A}{z_n},
\ \ i,j=1,\dots6\,\ ,\  A,B,m,n=1,2\nn,
\end{eqnarray}
since all of the remaining projection coefficients are determined by modular
invariance \cite{fff1,fff2}.
Each of the 55 independent coefficients can take two discrete
values $\pm1$ and thus a simple counting gives $2^{55}$
(that is approximately $10^{16.6}$) distinct models in the
class of superstring vacua under consideration.

The vector bosons from the untwisted sector generate an
$
SO(10)\times{U(1)}^3\times{SO(8)}^2
$
gauge symmetry.
Depending on the  choices of the projection coefficients,
extra gauge bosons may arise from
\beq
x=1+S+\sum_{i=1}^{6}e_i+z_1+z_2=\{{\bar{\eta}^{123},\bar{\psi}^{12345}}\}~,
\label{xvector}
\eeq
which enhances the gauge group $SO(10)\times{U(1)}\to E_6$.
Additional gauge bosons can arise as well from the sectors
$z_1,z_2$ and $z_1+z_2$ and enhance the hidden gauge group
${SO(8)}^2\to SO(16)$ or even ${SO(8)}^2\to E_8$.
Indeed, as was shown in ref. \cite{fknr},
for particular choices of the projection coefficients
a variety of gauge groups is obtained.
The classification in this paper is restricted to the case
in which all the gauge bosons from the sectors
$x$, $z_1$ ,$z_2$ and $z_1+z_2$ are projected out. Hence, in the entire
space of vacua the four dimensional gauge group is
$SO(10)\times {U(1)}^3\times{SO(8)}^2.
$

The matter spectrum from the untwisted sector is common to all models
and consists of six vectors of $SO(10)$ and 12 states that singlets under
the non-Abelian gauge groups.
The chiral spinorial representations arise necessarily from the following 48
twisted sectors
\begin{eqnarray}
B_{\ell_3^1\ell_4^1\ell_5^1\ell_6^1}^1&=&S+b_1+\ell_3^1 e_3+\ell_4^1 e_4 +
\ell_5^1 e_5 + \ell_6^1 e_6 \nn\\
B_{\ell_1^2\ell_2^2\ell_5^2\ell_6^2}^2&=&S+b_2+\ell_1^2 e_1+\ell_2^2 e_2 +
\ell_5^2 e_5 + \ell_6^2 e_6 \label{ss}\\
B_{\ell_1^3\ell_2^3\ell_3^3\ell_4^3}^3&=&
S+b_3+ \ell_1^3 e_1+\ell_2^3 e_2 +\ell_3^3 e_3+ \ell_4^3 e_4\nn
\end{eqnarray}
where $\ell_i^j=0,1$;
$b_3=b_1+b_2+x=1+S+b_1+b_2+\sum_{i=1}^6 e_i+\sum_{n=1}^2 z_n$ and $x$ is
given in eq. (\ref{xvector}).

The states that arise from the sectors in (\ref{ss})
are  spinorials of $SO(10)$ and one can obtain at most one
spinorial ($\bf 16$ or
$\bf {\overline{{16}}}$) per sector and thus totally 48 spinorials.
The states in the vector representation of $SO(10)$
arise necessarily from the $x$--mapped twisted sectors
$B_{\ell_3^i\ell_4^i\ell_5^i\ell_6^i}^i+x$\ ,  $(i=1,2,3)$, accompanied
always by six singlets under $SO(10)\times SO(8)\times SO(8)$.

The string vacua generically may contain additional hidden matter states
that transform under the hidden sector gauge group. These arise generically
from the sectors
$B_{\ell_3^i\ell_4^i\ell_5^i\ell_6^i}^i+x$\ ,
$B_{\ell_3^i\ell_4^i\ell_5^i\ell_6^i}^i+x+z_1$\ , and
$B_{\ell_3^i\ell_4^i\ell_5^i\ell_6^i}^i+x+z_2$\ ,  where $(i=1,2,3)$.
The hidden sector matter states appear in general in vector representations,
and may be chiral with respect to the unbroken $U(1)$ symmetries, defined
by the ${\bar\eta}_1$, ${\bar\eta}_2$ and ${\bar\eta}_3$ world--sheet fermions.
Our analysis here focuses on the observable sector states and
neglects the hidden sector matter states. An investigation
that includes the hidden matter states is of interest, in particular
in regard to the modular properties of this space of vacua, but their
inclusion is left for future work.

This construction therefore separates the fixed points of the $Z_2\times
Z_2$ orbifold into different sectors. This enables the analysis of
the  GGSO projection on the spectrum from each individual fixed point
separately. Hence, depending on the choice of the GGSO projection
coefficients we can distinguish several possibilities for the
spectrum from each individual fixed point. For example,
in the case of enhancement
of the $SO(10)$ symmetry to $E_6$ each individual fixed point
gives rise to spinorial, as well as vectorial representation of
$SO(10)$, which are embedded in the $27$ representation of $E_6$.
When $E_6$ is broken each fixed point typically will give rise
to either spinorial or vectorial representation of $E_6$. However,
there exist also rare situations, depending on the choice of
GGSO phases,
where a fixed point can yield a spinorial as well as vectorial
representation of $SO(10)$ without enhancement. The crucial point,
however, is that the GGSO projections can be written as simple
algebraic conditions, and hence the classification is amenable
to a computerized analysis.

\section{Counting the twisted matter spectrum\label{count}}

The counting of spinorials proceeds as follows.
For each $SO(10)$ spinorial from a twisted sector $B^i_{pqrs}$
defined in (\ref{ss})
we can write down the associated projector
$P^i_{pqrs}=0,1$, in terms of the GGSO projection coefficients.
The explicit expressions for the 48 projectors are
\begin{eqnarray}
P_{p^1q^1r^1s^1}^{(1)}&=&
\frac{1}{4}\,\left(1-\cc{e_1}{B_{p^1q^1r^1s^1}^{(1)}}\right)\,
\cdot\left(1-\cc{e_2}{B_{p^1q^1r^1s^1}^{(1)}}\right)\,\nn\\
&&\cdot\frac{1}{4}\left(1-\cc{z_1}{B_{p^1q^1r^1s^1}^{(1)}}\right)\,
\cdot\left(1-\cc{z_2}{B_{p^1q^1r^1s^1}^{(1)}}\right)\nn\\
P_{p^2q^2r^2s^2}^{(2)}&=&
\frac{1}{4}\,\left(1-\cc{e_3}{B_{p^2q^2r^2s^2}^{(2)}}\right)\,
\cdot\left(1-\cc{e_4}{B_{p^2q^2r^2s^2}^{(2)}}\right)\,\nn\\
&&\cdot\frac{1}{4}\,\left(1-\cc{z_1}{B_{p^2q^2r^2s^2}^{(2)}}\right)\,
\cdot\left(1-\cc{z_2}{B_{p^2q^2r^2s^2}^{(2)}}\right)\label{proj}
\\
P_{p^3q^3r^3s^3}^{(3)}&=&
\frac{1}{4}\,\left(1-\cc{e_5}{B_{p^3q^3r^3s^3}^{(3)}}\right)\,
\cdot\left(1-\cc{e_6}{B_{p^3q^3r^3s^3}^{(3)}}\right)\,\nn\\
&&\cdot\frac{1}{4}\,\left(1-\cc{z_1}{B_{p^3q^3r^3s^3}^{(3)}}\right)\,
\cdot\left(1-\cc{z_2}{B_{p^3q^3r^3s^3}^{(3)}}\right)~.\nn
\end{eqnarray}
When $P^i_{pqrs}=1$  there is a surviving spinorial
(${\bf 16}$ or ${\bf \overline{16}}\,$).
For the surviving spinorial ($P^i_{pqrs}=1$)
the  chirality (${\bf 16}$ or ${\bf \overline{16}}\,$)
is determined from the associated chirality
coefficient $X^i_{pqrs}=\pm1$, where
\begin{eqnarray}
X_{p^1q^1r^1s^1}^{(1)}&=&\cc{S+b_2+
(1-r^1) e_5+(1-s^1) e_6}{B_{p^1q^1r^1s^1}^{(1)}}
\nn
\\
X_{p^2q^2r^2s^2}^{(2)}&=&\cc{S+b_1+
(1-r^2) e_5+(1-s^2) e_6}{B_{p^2q^2r^2s^2}^{(2)}}
\\
X_{p^3q^3r^3s^3}^{(3)}&=&\cc{S+b_1+(1-r^3) e_3+
(1-s^3) e_4}{B_{p^3q^3r^3s^3}^{(3)}}\nn\\
&=&\cc{S+b_2+(1-p^3) e_1+
(1-q^3) e_2}{B_{p^3q^3r^3s^3}^{(3)}}~.
\nn
\end{eqnarray}
These formulas are dictated by the vector intersections
\begin{eqnarray}
 S+b_2+(1-r^1) e_5+(1-s^1) e_6 \cap B_{p^1q^1r^1s^1}^{(1)} &=& \nn\\
S+b_1+(1-r^2) e_5+(1-s^2) e_6 \cap B_{p^2q^2r^2s^2}^{(2)} &=& \nn\\
S+b_1+(1-r^3) e_3+ (1-s^3) e_4 \cap B_{p^3q^3r^3s^3}^{(3)} &=& \nn\\
S+b_2+(1-p^3) e_1+ (1-q^3) e_2 \cap B_{p^3q^3r^3s^3}^{(3)}
&=&\{\bar{\psi}^{12345}\}~.\nn
\end{eqnarray}

Using the above results, we can easily calculate the number of
spinorials/antispinorial per sector
\begin{eqnarray}
S_{\pm}^{(i)}&=&\sum_{pqrs}
\frac{1\pm X^{(i)}_{p^iq^ir^is^i}}{2} P_{p^iq^ir^is^i}^{(i)}\ ,
\ i=1,2,3\label{cc}
\end{eqnarray}

The counting of $SO(10)$ vectorials can proceed in a similar way.
For each vectorial generating sector
$(B_{pqrs})^i+x$ the associated projector  $(\tilde{P}_{pqrs})^i$ is
obtained from (\ref{proj})
using the replacement $(B_{pqrs})^i\to (B_{pqrs})^i+x$.
Since there is no chirality in this case the number of
vectorials per sector is just the sum of the projectors
\begin{eqnarray}
V^{(i)}=\sum_{pqrs} (\tilde{P}_{pqrs})^{(i)}
\end{eqnarray}

The total number of vectors ($V$), the total number of spinors plus
anti--spinors ($S_t$), and the net number of spinors minus anti--spinors
($S_c$) are given by
\begin{eqnarray}
V=\sum_{i=1}^3 V^{(i)},
\end{eqnarray}
\begin{eqnarray}
S_t=\sum_{i=1}^3 S_{+}^{(i)}+S_{-}^{(i)}
\end{eqnarray}
and
\begin{eqnarray}
S_c=\sum_{i=1}^3 S_{+}^{(i)}-S_{-}^{(i)},
\end{eqnarray}
respectively.

The mixed projection coefficients entering the above formulas can be
decomposed in terms of the
independent phases $\cc{v_i}{v_j}, i>j$.
After some algebra we come to the conclusion that for the counting of the
spinorial/antispinorial and
vectorial $SO(10)$ states the phases
$\cc{e_i}{e_i}, i=1,\dots,6$,
$\cc{z_A}{z_A}, A=1,\dots,2$,
$\cc{b_I}{b_I}, I=1,\dots,2$ as well as
$\cc{e_3}{b_1}, \cc{e_4}{b_1}, \cc{e_1}{b_2}, \cc{e_2}{b_2}$
are not relevant. Moreover the phase $\cc{b_1}{b_2}$ is related to the
total chirality flip.
This leaves a set of 40 independent phases which is still too
large for a manageable computer analysis.

To reduce the number of independent GGSO phases further,
we restrict the classification to the space of models in which
the four dimensional gauge group arises solely from the untwisted
sector. This fixes some additional phases, as we detail below.
With respect to this subclass of four dimensional solutions
the classification is complete.

We can get more information regarding the possible spinorial and
vectorial multiplicities
per plane by rewriting the projectors (\ref{proj}) in the form of a
system of equations.
Introducing the notation
\ba
\cc{a_i}{a_j}=e^{i \pi \oo{a_i}{a_j}}\,\ ,\  \oo{a_i}{a_j}=0,1
\ea
with the properties
\ba
\oo{a_i}{a_j+a_k}&=&\oo{a_i}{a_j}+\oo{a_i}{a_k} \ ,
\forall\ a_{i}: \{\psi^\mu\}\cap a_i=\O
\\
\oo{a_i}{a_j}&=&\oo{a_j}{a_i} \ ,
\forall\ a_{i},a_{j}: a_i\cdot a_j=0\ {\rm mod}\ 4
\ea
where $\# (a_i\cdot a_j)\equiv\# \left[a_i\cup a_j - a_i\cap a_j\right]$.
The projectors can be written as system of equations (one per plane)
\ba
\Delta^{(I)}\,U_{16}^{(I)}=Y_{16}^{(I)}\ ,\
\Delta^{(I)}\,U_{10}^{(I)}=Y_{10}^{(I)}\ ,\ I=1,2,3
\label{sysi}
\ea
where the unknowns are the fixed point labels
\ba
U_{16}^{(I)}=
\left[
\begin{array}{c}
p_{16}^I\\
q_{16}^I\\
r_{16}^I\\
s_{16}^I
\end{array}
\right]
\ ,\
U_{10}^{(I)}=
\left[
\begin{array}{c}
p_{10}^I\\
q_{10}^I\\
r_{10}^I\\
s_{10}^I
\end{array}
\right]
\ea
and
\ba
\Delta^{(1)}&=&\left[
\begin{array}{cccc}
\oo{e_1}{e_3}&\oo{e_1}{e_4}&\oo{e_1}{e_5}&\oo{e_1}{e_6}\\
\oo{e_2}{e_3}&\oo{e_2}{e_4}&\oo{e_2}{e_5}&\oo{e_2}{e_6}\\
\oo{z_1}{e_3}&\oo{z_1}{e_4}&\oo{z_1}{e_5}&\oo{z_1}{e_6}\\
\oo{z_2}{e_3}&\oo{z_2}{e_4}&\oo{z_2}{e_5}&\oo{z_2}{e_6}
\end{array}
\right]
\nn
\\
\Delta^{(2)}&=&\left[
\begin{array}{cccc}
\oo{e_3}{e_1}&\oo{e_3}{e_2}&\oo{e_3}{e_5}&\oo{e_3}{e_6}\\
\oo{e_4}{e_1}&\oo{e_4}{e_2}&\oo{e_4}{e_5}&\oo{e_4}{e_6}\\
\oo{z_1}{e_1}&\oo{z_1}{e_2}&\oo{z_1}{e_5}&\oo{z_1}{e_6}\\
\oo{z_2}{e_1}&\oo{z_2}{e_2}&\oo{z_2}{e_5}&\oo{z_2}{e_6}
\end{array}
\right]
\label{deltai}
\\
\Delta^{(3)}&=&\left[
\begin{array}{cccc}
\oo{e_5}{e_1}&\oo{e_5}{e_2}&\oo{e_5}{e_3}&\oo{e_5}{e_4}\\
\oo{e_6}{e_1}&\oo{e_6}{e_2}&\oo{e_6}{e_3}&\oo{e_6}{e_4}\\
\oo{z_1}{e_1}&\oo{z_1}{e_2}&\oo{z_1}{e_3}&\oo{z_1}{e_4}\\
\oo{z_2}{e_1}&\oo{z_2}{e_2}&\oo{z_2}{e_3}&\oo{z_2}{e_4}
\end{array}
\right]
\nn
\ea
\ba
Y_{16}^{(1)}=
\left[
\begin{array}{c}
\oo{e_1}{b_1}\\
\oo{e_2}{b_1}\\
\oo{z_1}{b_1}\\
\oo{z_2}{b_1}
\end{array}
\right]
\ \
\ ,\
Y_{16}^{(2)}=
\left[
\begin{array}{c}
\oo{e_3}{b_2}\\
\oo{e_4}{b_2}\\
\oo{z_1}{b_2}\\
\oo{z_2}{b_2}
\end{array}
\right]
\ ,\
Y_{16}^{(3)}=
\left[
\begin{array}{c}
\oo{e_5}{b_3}\\
\oo{e_6}{b_3}\\
\oo{z_1}{b_3}\\
\oo{z_2}{b_3}
\end{array}
\right]
\ea
\ba
Y_{10}^{(1)}=
\left[
\begin{array}{c}
\oo{e_1}{b_1+x}\\
\oo{e_2}{b_1+x}\\
\oo{z_1}{b_1+x}\\
\oo{z_2}{b_1+x}
\end{array}
\right]
\ \
\ ,\
Y_{10}^{(2)}=
\left[
\begin{array}{c}
\oo{e_3}{b_2+x}\\
\oo{e_4}{b_2+x}\\
\oo{z_1}{b_2+x}\\
\oo{z_2}{b_2+x}
\end{array}
\right]
\ ,\
Y_{10}^{(3)}=
\left[
\begin{array}{c}
\oo{e_5}{b_3+x}\\
\oo{e_6}{b_3+x}\\
\oo{z_1}{b_3+x}\\
\oo{z_2}{b_3+x}
\end{array}
\right]
\ea

Using standard linear algebra results,
we find that the systems of equations (\ref{sysi}) have solutions
when the
rank of the matrix $\Delta^{(I)}$ equals to the rank of the associated
augmented matrices: $[\Delta^{(I)},Y_{16}^{(I)}]$
for spinorials and $[\Delta^{(I)},Y_{10}^{(I)}]$ for vectorials.
In our case the number of solutions and thus the
total number of spinorials and vectorials per orbifold plane are given by
\ba
S^{(I)}=\left\{
\begin{array}{ll}
2^{4-{\rm rank}\left(\Delta^{(I)}\right)}
&, {\rm rank}\left(\Delta^{(I)}\right)=
{\rm rank}\left[\Delta^{(I)},Y_{16}^{(I)}\right]\\
 & \\
0&, {\rm rank}\left(\Delta^{(I)}\right)<{\rm rank}
\left[\Delta^{(I)},Y_{16}^{(I)}\right]
\end{array}
\right.
\ ,\ I=1,2,3
\label{ssi}
\ea
\ba
V^{(I)}=\left\{
\begin{array}{ll}
2^{4-{\rm rank}\left(\Delta^{(I)}\right)}&, {\rm rank}
\left(\Delta^{(I)}\right)={\rm
rank}\left[\Delta^{(I)},Y_{10}^{(I)}\right]\\
 & \\
0&, {\rm rank}\left(\Delta^{(I)}\right)<{\rm rank}
\left[\Delta^{(I)},Y_{10}^{(I)}\right]
\end{array}
\right.
\ ,\ I=1,2,3
\label{vvi}
\ea
 The results of the application of formulas (\ref{ssi}),
(\ref{vvi}) are presented in Table \ref{tsv}.
\begin{table}[!h]
\begin{center}
\begin{tabular}{|c|c|c|c|c|}
\hline
rank
$\left(\Delta^{(I)}\right)$&rank$\left[\Delta^{(I)},Y_{16}^{(I)}\right]$
&rank $\left[\Delta^{(I)},Y_{10}^{(I)}\right]$
&\# of Spinorials&\# of vectorials\\
\hline
4&4&4&1&1\\
\hline
3&4&4&0&0\\
 &3&4&2&0\\
 &4&3&0&2\\
 &3&3&2&2\\
 \hline
2&3&3&0&0\\
 &2&3&4&0\\
 &3&2&0&4\\
 &3&3&4&4\\
 \hline
1&2&2&0&0\\
 &1&2&8&0\\
 &2&1&0&8\\
 &1&1&8&8\\
 \hline
0&1&1&0&0\\
 &0&1&16&0\\
 &1&0&0&16\\
 &0&0&16&16\\
 \hline
\end{tabular}
\end{center}
\caption{\label{tsv}\it
Total number of $SO(10)$ spinorial and vectorial representations
in a given orbifold plane $I=1,2,3$ for all possible ranks of the
projection matrices
$\left(\Delta^{(I)}\right)$,
$\left[\Delta^{(I)},Y_{16}^{(I)}\right]$, and
$\left[\Delta^{(I)},Y_{10}^{(I)}\right]$.}
\end{table}
\section{The four dimensional gauge group}\label{4dgroup}
For all models  generated by the basis set (\ref{basis})
the gauge bosons of the null sector give rise to
a gauge symmetry
\ba
G=SO(10)\times U(1)^3\times{SO(8)}_1\times{SO(8)}_2\ .
\label{mg}
\ea
Additional gauge bosons may arise
 from the sectors
$$
x\ ,\ z_1\ ,\ z_2\ ,\ z_1+z_2
$$
that  can lead to enhancements of the observable and/or the hidden gauge group.
These enhancements are model dependent, and hence depend on specific choices
of GGSO phases. These enhancements include:
\newcounter{mlls}
\begin{list}{(\Roman{mlls})}{\usecounter{mlls}}
\item The $x$--sector gauge bosons give rise to
$SO(10)\times{U(1)}\to{E_6}$ enhancement when
\ba
\oo{e_i}{x}=\oo{z_k}{x}=0\ \forall\ i=1,\dots,6\ ,\ k=1,2~.
\label{eixr}
\ea
\item The $z_1+z_2$--sector
gauge bosons can lead to ${SO(8)}^2\to{SO(16)}$
enhancement when
\ba
\oo{e_i}{z_1}=\oo{e_i}{z_2}\ \forall\ i=1,\dots,6\ ,\
\oo{b_m}{z_1}=\oo{b_m}{z_2}\ \forall\ m=1,2
\label{eiz1bmz1}
\ea
\item The $z_k$--sectors, $(k=1,2)$,
enhancements involve right--moving fermionic
oscillators and belong in two classes depending on the
value of $\oo{z_1}{z_2}$:

(a) for $\oo{z_1}{z_2}=1$ we obtain gauge bosons that
involve $z_1$ and/or $z_2$ oscillators, namely
$\{\bar{\phi}^{1\dots8}\}$. These lead
to hidden group enhancements, and particularly to
${SO(8)}^2\to{SO(16)}$ when
\ba
\oo{z_1}{z_2}=1\ ,\ \oo{e_i}{z_k}=\oo{b_m}{z_k}=0\ \forall\ i=1,\dots,6
\label{z1z2eizk}
\ea
(b) for $\oo{z_1}{z_2}=0$ we obtain gauge bosons that involve
oscillators not included in $z_1,z_2$ and lead
thus to gauge bosons that mix
${SO(8)}_1$ or/and  ${SO(8)}_2$  with other  group factors in
(\ref{mg}). These include
\ba
\oo{z_1}{z_2}=0\ ,\ \oo{e_i}{z_k}=0\ \forall\ i=1,\dots,6
\label{z1z2eizk2}
\ea
for $k=1$ or/and $k=2$. In this case the gauge group enhancement
includes several possibilities,
depending on the $\oo{b_m}{z_k}$ we can obtain\\

$SO(10)\times{SO(8)}_k\to{SO(18)}$,\\
${SO(8)}_k\times{U(1)}\to{SO(10)}$,\\
or
${SO(8)}^2\times{U(1)}^2\to{SO(10)}^2$.

Moreover for $\oo{z_1}{z_2}=0$ and particular choices
of $\oo{e_i}{z_k}$ and $\oo{b_m}{z_k}$ we can have
${SO(8)}_k\to{SO(9)}$ enhancements.\\
\end{list}
Mixed combinations of the above are possible when the
conditions on the associated GGSO coefficients are
compatible. For example combination of gauge bosons (II)
with those in (IIIb) can lead to
$SO(10)\times{SO(8)}^2\to{SO(26)}$ enhancement.

In the present work
we restrict to models where all the additional gauge bosons from the
sectors $x$, $z_1+z_2$ and $z_k$ sectors
are absent.
This is achieved for appropriate choice of the GSO phases
such that the above requirements (\ref{eixr}--\ref{z1z2eizk2})
are not satisfied.

\section{Results}\label{results}

\begin{figure}[!h]
\centerline{\epsfxsize 5.0 truein \epsfbox{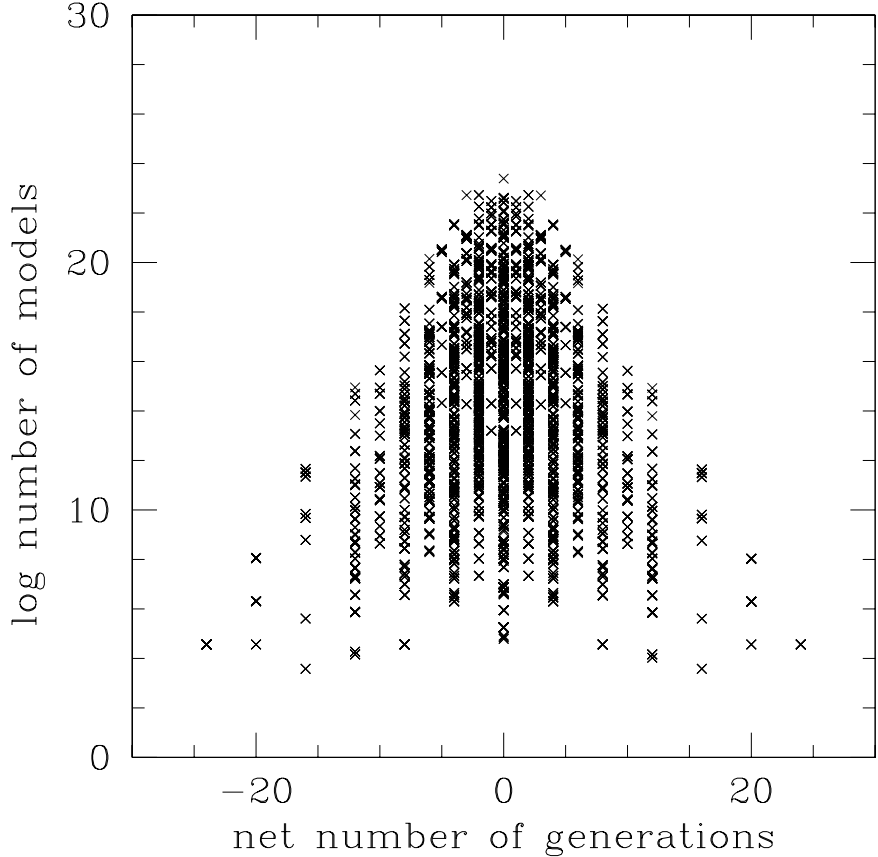}}
\caption[1]{\label{lscatnetgenall}
Scatter plot of the logarithm of the number of models versus the
net number of chiral families, $S_c$.}
\end{figure}

\begin{figure}[!h]
\centerline{\epsfxsize 5.0 truein \epsfbox{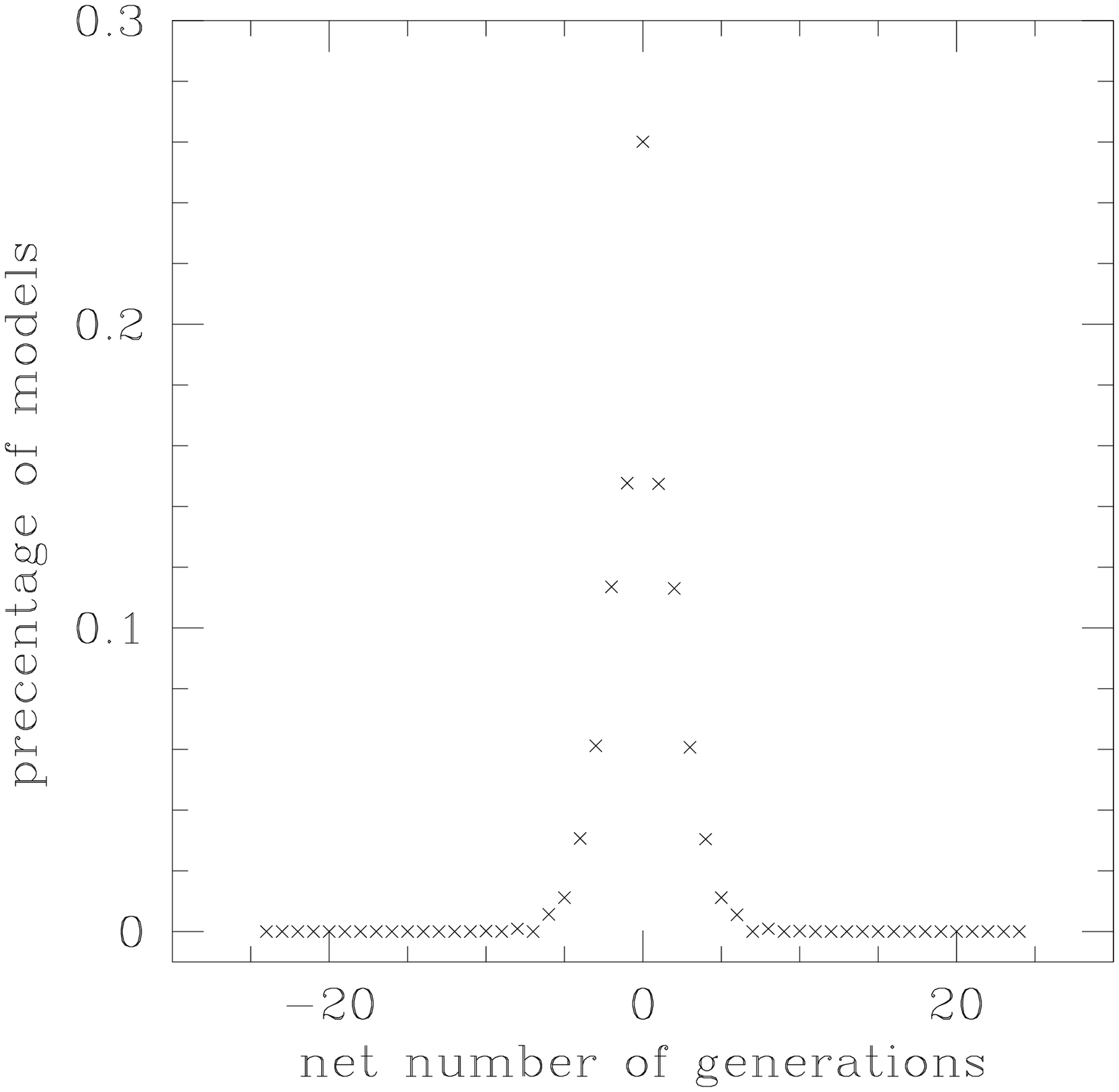}}
\caption[1]{\label{precentall}
Percentage of models with a net number of generations, $S_c$.}
\end{figure}

We classify the string vacua, under the no--enhancement restrictions
described above, according to the numbers of spinors $S_+$, anti--spinors
$S_-$ and vectors $V$.
The results of this classification
does not differ substantially
from the random model generation search that was done in ref. \cite{fkr}.
The distribution of the models with respect to the net number
of chiral families $S_c=S_+-S_-$,
and the percentage of models with a given
$S_c$, are displayed in figures \ref{lscatnetgenall}
and \ref{precentall}, respectively,
and can be compared with the corresponding figures
in ref. \cite{fkr}. The new figures are denser and represents a
scan of a larger set of models, but their qualitative appearance
is similar to those generated by the statistical analysis
of ref. \cite{fkr}.

A statistical analysis approach to the study
of string vacua has been of contemporary interest
\cite{statistical}. Our results
in this respect may be viewed as providing encouragement that
the statistical analysis approach may indeed provide some insight
into the properties of large classes of string compactifications.
As in ref. \cite{fkr} we observe a bell shape distribution
that peaks for vanishing net number of generations.
Similarly to ref. \cite{fkr} models with a net number
of 7, 9 11, 13, 14, 15, 17, 18, 19, 21, 22, 23 chiral generations
are not found in the distribution.
The results of the statistical analysis of ref. \cite{fkr}
conquer with the complete method of classification of the current
analysis.

The distribution exhibits a symmetry with respect to the exchange
of spinor $S_+$ and anti--spinors $S_-$.
The symmetry is not identical to the mirror symmetry on Calabi--Yau
manifolds with $E_6$ symmetry \cite{vafawitten}.
Indeed in our models
there exist an overall chirality phase $\cc{b_1}{b_2}=\pm1$, which is
fixed in our analysis.
This overall chirality phase corresponds to the discrete torsion in
the $N=1$ partition function and fixes the overall chirality of
the models. The change of this phase, according to some
arguments in the literature \cite{vafawitten}, corresponds to
the mirror symmetry transformation on Calabi--Yau manifolds with
$E_6$ symmetry and $(2,2)$ superconformal compactifications.
The $S_+\leftrightarrow S_-$ exchange symmetry
of the vacua that we classify in this work corresponds necessarily to
a new mirror like symmetry, which is independent of the discrete torsion
associated with the overall chirality phase $\cc{b_1}{b_2}$.
Further studies of the origin of the new mirror like symmetry
arising naturally in $(2,0)$ superconformal compactifications
related to the vacua examined here will be reported in future
publications.

\begin{figure}[!h]
\centerline{\epsfxsize 5.0 truein \epsfbox{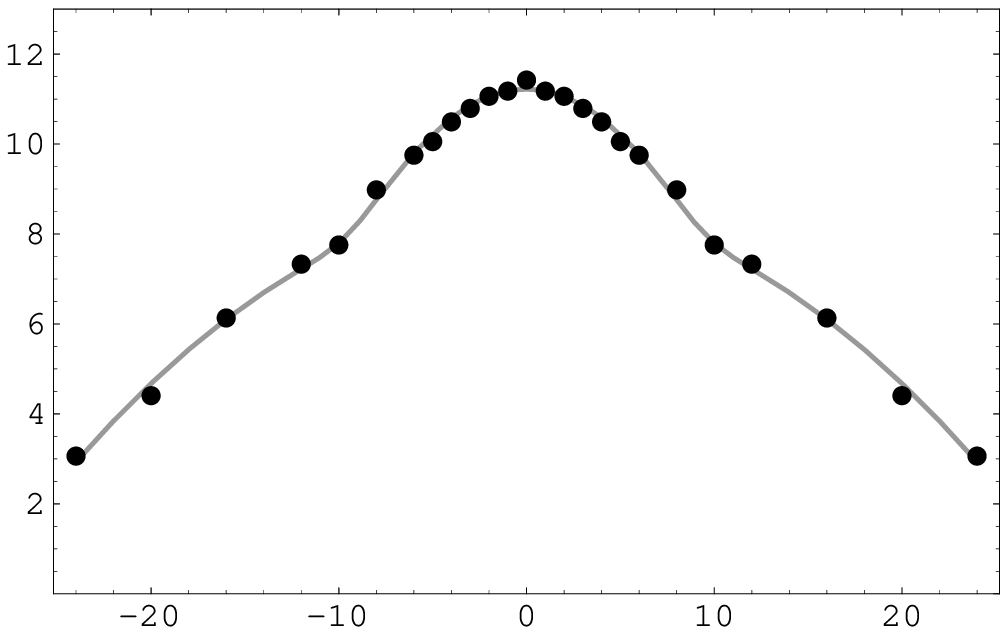}}
\caption[1]{\label{gaussianfit}
Total number of models as a function of net chirality.
The gray line corresponds to the
sum of Gaussians $f=A\,e^{-\alpha x^2}+B\,e^{-\alpha x^2/4}$
where $A=1.64\times 10^{11}$, $B=4.39\times 10^8$ and
$\alpha=9.13\times 10^{-2}$.}
\end{figure}

In figure \ref{gaussianfit} we demonstrate that the
distribution of the number of models
as a function of the net number of chiral families in not well fitted
with a Gaussian curve, as suggested in ref. \cite{douglas}.
The distribution is fitted better
with a sum of two Gaussian functions as illustrated
in figure \ref{gaussianfit}.

More interestingly we find that
the space of $Z_2\times Z_2$ orbifold models exhibits a novel symmetry
under the exchange of the total number of
vectorial representations $V$ and the total number of
spinorial plus anti--spinorial
representations $S_t$. Thus, for any given model with a total number of
$SO(10)$ $V$--representations, there exist a corresponding model in which
the total number of $SO(10)$ $S_t$--representations
is the same. Below we turn to investigate this symmetry in some detail.

\section{Spinor--vector duality}\label{svduality}

The existence of a $V\leftrightarrow S_t$ duality exchange symmetry
is apparent when the $SO(10)$ symmetry is enhanced
to $E_6$. In this case the chiral matter states arise from the
$27$ and $\overline{27}$ representations of $E_6$, which decompose
under $SO(10)$ as
\beqn
27 & \equiv & 16 \oplus 10 \oplus 1 \nonumber\\
\overline{27} & \equiv & \overline{16} \oplus 10 \oplus 1
\label{e6toso10}
\eeqn
{}From eq. (\ref{e6toso10}) it is seen that in this case the total number
of spinorial $16\oplus{\overline 16}$ $SO(10)$ representations is equal to the
total number of vectorial $10$ representations, and such models are
self--dual under the exchange. Thus, $V\leftrightarrow S_t$ duality
is trivial in the case of $(2,2)$ Calabi--Yau compactifications.
However, over the space of $(2,0)$ vacua that we scan in this work,
the $SO(10)$ symmetry is not enhanced to
$E_6$ symmetry.
Nevertheless, the distribution of vacua still exhibits this symmetry.
Furthermore, we find that the
$V\leftrightarrow S_t$ duality holds separately for each twisted plane.
In table \ref{tableofexamples} we illustrate the $V\leftrightarrow S_t$
duality on the first plane. The duality is observed by noting that
for a fixed number of representations, summing over the number
of models with a total number of spinor plus anti--spinor
representations
produces the identical number of models with the same number
of vector representations. Thus, for example, summing over the
number of models in the first three rows produces the number
of models in the fourth row. Considering that the integral numbers
involved are quite high, the resulting equalities are quite astounding!
\begin{table}[!h]
\begin{center}
\begin{tabular}{|ccc|ccc|ccc|c|}
\hline
      & First Plane & &    & Second plane & & & Third Plane & & \\
\hline
$s$ &${\bar s}$& $v$ &$s$&${\bar s}$&$v$&$s$&${\bar s}$&$v$& \# of models \\
\hline
2 & 0 & 0 &    0 & 0 & 0 &    0 & 0 & 0 & 1325963712 \\
0 & 2 & 0 &    0 & 0 & 0 &    0 & 0 & 0 & 1340075584 \\
1 & 1 & 0 &    0 & 0 & 0 &    0 & 0 & 0 & 3718991872 \\
\hline
0 & 0 & 2 &    0 & 0 & 0 &    0 & 0 & 0 & 6385031168 \\
\hline
\hline
4 & 0 & 0 &    0 & 0 & 0 &    0 & 0 & 0 & ~111944544  \\
3 & 1 & 0 &    0 & 0 & 0 &    0 & 0 & 0 & ~250947136  \\
2 & 2 & 0 &    0 & 0 & 0 &    0 & 0 & 0 & 1059624448  \\
1 & 3 & 0 &    0 & 0 & 0 &    0 & 0 & 0 & ~251936192  \\
0 & 4 & 0 &    0 & 0 & 0 &    0 & 0 & 0 & ~113437024  \\
\hline
0 & 0 & 4 &    0 & 0 & 0 &    0 & 0 & 0 & 1787889344 \\
\hline
\hline
0 & 8 & 0 &    0 & 0 & 0 &    0 & 0 & 0 & ~~~~535280  \\
2 & 6 & 0 &    0 & 0 & 0 &    0 & 0 & 0 & ~~~8084480  \\
4 & 4 & 0 &    0 & 0 & 0 &    0 & 0 & 0 & ~~34050304  \\
6 & 2 & 0 &    0 & 0 & 0 &    0 & 0 & 0 & ~~~8053760  \\
8 & 0 & 0 &    0 & 0 & 0 &    0 & 0 & 0 & ~~~~529040  \\
\hline
0 & 0 & 8 &    0 & 0 & 0 &    0 & 0 & 0 & ~~51252864  \\
\hline
\hline
0 &16 & 0 &    0 & 0 & 0 &    0 & 0 & 0 & ~~~~~~~272  \\
4 &12 & 0 &    0 & 0 & 0 &    0 & 0 & 0 & ~~~~~~9792  \\
6 &10 & 0 &    0 & 0 & 0 &    0 & 0 & 0 & ~~~~~26112  \\
8 & 8 & 0 &    0 & 0 & 0 &    0 & 0 & 0 & ~~~~~84000  \\
10& 6 & 0 &    0 & 0 & 0 &    0 & 0 & 0 & ~~~~~26112  \\
12& 4 & 0 &    0 & 0 & 0 &    0 & 0 & 0 & ~~~~~~9792  \\
16& 0 & 0 &    0 & 0 & 0 &    0 & 0 & 0 & ~~~~~~~272  \\
\hline
0 & 0 & 16 &    0 & 0 & 0 &    0 & 0 & 0 & ~~~156352  \\
\hline
\hline
\end{tabular}
\end{center}
\caption{\label{tableofexamples}\it
Examples of spinor--vector duality on the first twisted plane.
The total number of models with a given number of spinors plus antispinors
is equal to the total number of models with the same number
of vectors. The total number of models with a given number of spinors plus
antispinors is obtained by summing over the different distributions
of spinors and antispinors. }
\end{table}


\begin{figure}[!h]
\centerline{\epsfxsize 5.0 truein \epsfbox{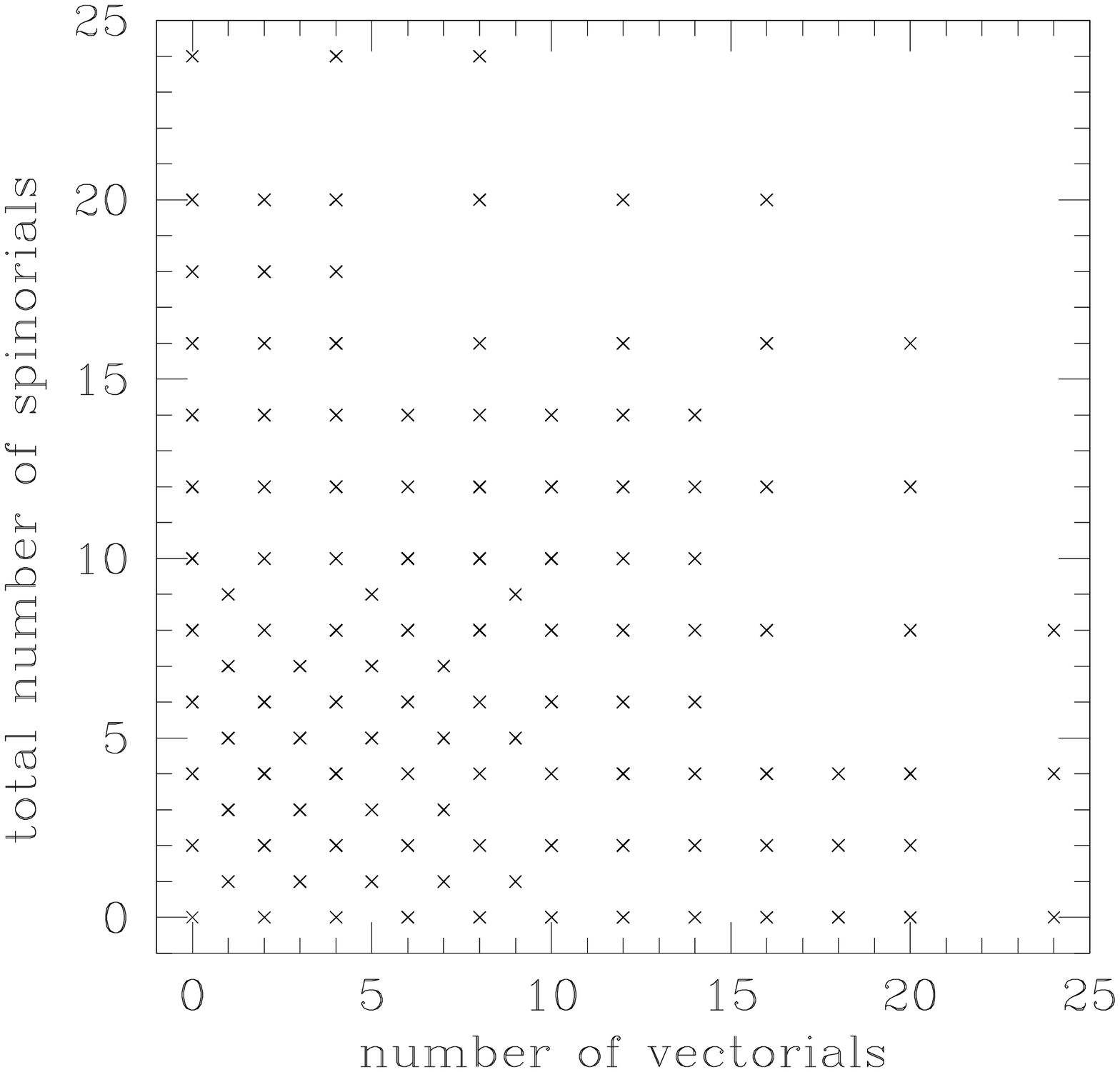}}
\caption[1]{\label{spvsvcall}
Total number of spinors plus antispinors, $S_t$,
versus the number of vectors, $V$,
occurring in the scanned space of vacua.
}
\end{figure}

In
figure \ref{spvsvcall} we display the total number
of spinors plus anti--spinors versus the number of vectors
occurring in the scan.
The figure is clearly symmetric under the exchange of the two axis,
which illustrates that for any model with a given number of spinors,
anti--spinors
and vectors, there is a corresponding model in which the number of
vectors is swapped
with the number of spinors plus anti--spinors.

\begin{figure}[!h]
\centerline{\epsfxsize 5.0 truein \epsfbox{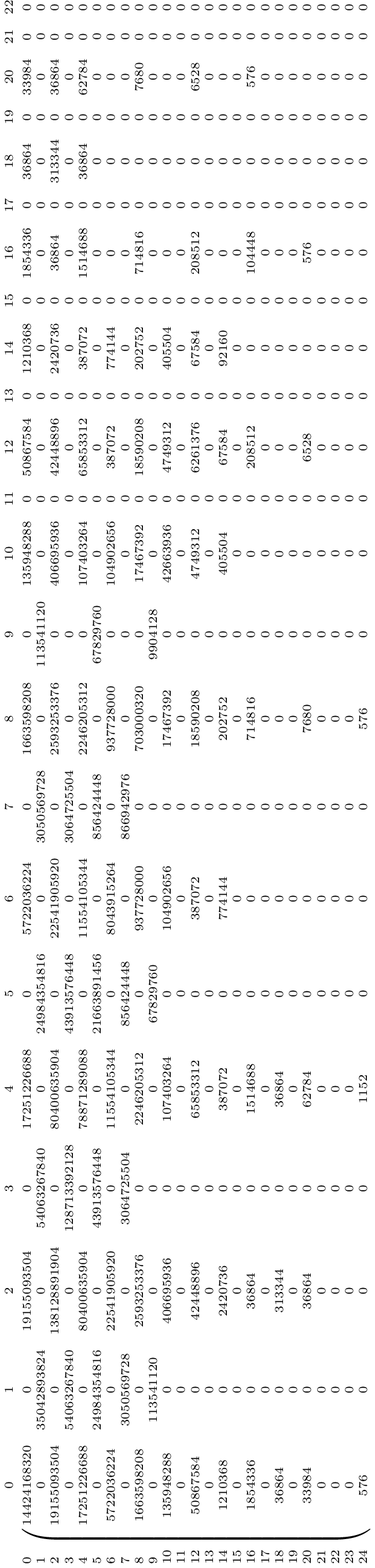}}
\caption[1]{\label{vectorspinormatrix}
Matrix of spinor plus antispinor versus vectors. The entries of the matrix
are the total number of models with a given number of spinors and vectors.
The symmetry of the matrix manifest the spinor--vector duality.}
\end{figure}

In eq. (\ref{vectorspinormatrix}) we display in a matrix form
the number of models for a given number of vectors
and spinors plus anti--spinors. The indices of the
raws and columns of the matrix
indicate the number of respective representations in the models,
whereas the entries are the number of models.
In each entry we sum over different configurations by which the
spinors, anti--spinor and vector representations are arranged
in the three twisted planes and fixed point sectors. Therefore,
the entries represent nontrivial sums over different configurations.
Examining the matrix in eq. (\ref{vectorspinormatrix})
it is seen that it a symmetric matrix reflecting the
invariance under exchange of vectors with spinors plus anti--spinors.

\begin{figure}[!h]
\centerline{\epsfxsize 5.5 truein \epsfbox{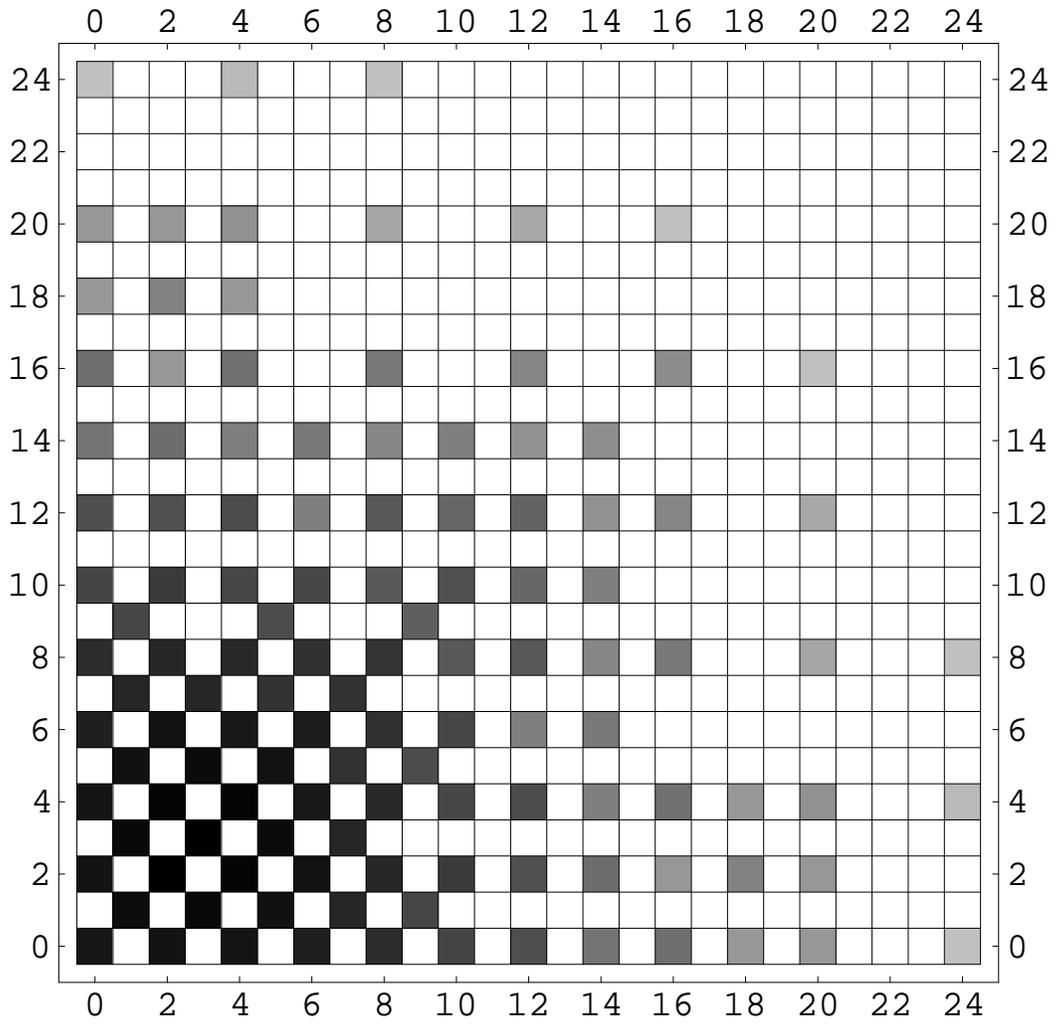}}
\caption[1]{\label{densityplot}
Density plot of the number of models versus the
number of vectors and spinors plus anti--spinors.}
\end{figure}

Figure \ref{densityplot}
is a graphic representation of eq. (\ref{vectorspinormatrix})
and shows a density plot of the number of models. The axis of the
plot are the number of vectors and the number of spinors plus
anti--spinors. The layout of the plot is similar to that
of figure \ref{spvsvcall}. The density of the number
of models, represented by the gray colouration,
exhibits the invariance under exchange of vectors
and spinors plus anti--spinors.

\vfill\eject
\vfill\eject

\vfill\eject

The $V\leftrightarrow S_t$ duality symmetry
reflects some modular properties
of the $N=1$ partition function.
This symmetry arises from a non--trivial discrete torsion induced
by reversing some of the GGSO projection coefficients.
To illustrate this discrete exchange we consider the simplified model
produced by the set of basis vectors

\begin{eqnarray}
v_1=1&=&\{\psi^\mu,\ \chi^{1,\dots,6},y^{1,\dots,6}, \omega^{1,\dots,6}| \nn\\
     & & ~~~\bar{y}^{1,\dots,6},\bar{\omega}^{1,\dots,6}, \bar{\eta}^{1,2,3},
        \bar{\psi}^{1,\dots,5},\bar{\phi}^{1,\dots,8}\},\nn\\
v_2=S&=&\{\psi^\mu,\chi^{1,\dots,6}\},\nn\\
v_{3}=z_1&=&\{\bar{\phi}^{1,\dots,4}\},\nn\\
v_{4}=z_2&=&\{\bar{\phi}^{5,\dots,8}\}.\nn\\
v_{5}=x&=&\{\bar{\psi}^{1,\dots,5},\bar{\eta}^{1,2,3}\},\nn\\
v_{6}=b_1&=&\{\chi^{34},\chi^{56},y^{34},y^{56}|\bar{y}^{34},
\bar{y}^{56},\bar{\eta}^1,\bar{\psi}^{1,\dots,5}\},\label{basis2}\\
v_{7}=b_2&=&\{\chi^{12},\chi^{56},y^{12},\omega^{56}|\bar{y}^{12},
\bar{\omega}^{56},\bar{\eta}^2,\bar{\psi}^{1,\dots,5}\}\nn
\end{eqnarray}

with the set of one--loop GGSO coefficients

\begin{equation}
{\bordermatrix{
              &  1&  S & & {z_1}&{z_2}&{x}& & {b_1}&{b_2}\cr
             1& -1&~~1 & & -1   &  -1 & -1  & &  -1     &  -1   \cr
             S&~~1&~~1 & & -1   &  -1 & -1  & & ~~1     & ~~1   \cr
          &   &    & &      &     &     & &         &       \cr
         {z_1}& -1& -1 & & -1   &  -1 & -1  & &  -1     &  -1   \cr
         {z_2}& -1& -1 & & -1   &  -1 &~~1  & & ~~1     & ~~1   \cr
           {x}& -1& -1 & & -1   & ~~1 &~~1  & &  -1     & ~~1   \cr
          &   &    & &      &     &     & &         &       \cr
         {b_1}& -1& -1 & & -1   & ~~1 &~~1  & &  -1     & ~~1   \cr
         {b_2}& -1& -1 & & -1   & ~~1 & -1  & & ~~1     &  -1   \cr}}
\label{freephases}
\end{equation}

The gauge group of this model is
$SO(12)\times SO(10)\times U(1)^3\times SO(8)\times SO(8)$ and
arises solely from the null sector. The gauge bosons arising
from all other sectors are projected out.
In this model the number of the $SO(10)$ spinorial representations
$\# S_+=8$ arising from the sector
$$b_3\equiv 1+z_1+z_2+b_1+b_2~.$$
There are no more spinorial representations from the
$b_1$ sector nor from the
$b_2$ sector. The sectors $b_3+x$ and $b_{1,2}+x+z_{1}$
each, produce eight multiplets that transform
in the 8 representation of the hidden $SO(8)_1$ gauge group,
whereas the sectors $b_{1,2,3}+x+z_2$
each, produce eight multiplets that transform
in the 8 representation of the hidden $SO(8)_2$ gauge group.
The sector $z_2$ produces a single state that transforms as
$8_{\rm v}\otimes 8_{\rm c}$ under $SO(8)_1\otimes SO(8)_2$.

Switching on a discrete torsion defined by the phase change
\beq
\cc{z_1}{b_1}=+1\rightarrow\cc{z_1}{b_1}=-1~,
\label{phasechange}
\eeq
the eight $SO(10)$ spinorial representations
from the sector
$b_3$ are now projected out, whereas the sector $b_3+x$ generates eight
vectorial 10 representation of $SO(10)$ plus additional $8\times6$ $SO(10)$
singlets. The remaining observable spectrum, which is charged under $SO(10)$,
is identical in these two models, with and without discrete torsion.
Hence, the discrete torsion defined in eq.
(\ref{phasechange}) induces the duality transformation that exchanges the
$SO(10)$ spinorial and vectorial representations.
Additionally, the hidden sector matter spectrum is also modified.
Indeed, in the presence of torsion
the sectors $b_2+x$ and $b_{1}+x+z_{2}$ each
produce eight multiplets that transform
in the 8 representation of the hidden $SO(8)_2$ gauge group,
whereas the sectors $b_{1,2}+x+z_1$
each, produce eight multiplets that transform
in the 8 representation of the hidden $SO(8)_1$ gauge group.
The sector $z_2$ produces a single state that transforms as
$8_{\rm v}\otimes 8_{\rm c}$ under $SO(8)_1\otimes SO(8)_2$.

In the next section
we present a general proof, based on the solutions of eqs. (\ref{sysi}),
for the $V\leftrightarrow S_t$ duality symmetry over the space of vacua.

\section{Analytic proof of spinor-vector duality}\label{proof}

In section \ref{count} a system of the projection equations
(\ref{sysi}--\ref{vvi}) that
determines the number of spinor and vector representations per twisted
plane in algebraic form. This involves the $4\times 4$ binary matrices
$\Delta^{(I)}$ of eq. (\ref{deltai}) and the augmented matrices
$\left[\Delta^{(I)},Y_{16}^{(I)}\right]$ and
$\left[\Delta^{(I)},Y_{10}^{(I)}\right]$.
Since the $V\leftrightarrow S_t$ duality interchanges
spinors and vectors and since the number of the spinor and vector
states relates to the rank of these matrices, as given in eqs.
(\ref{ssi}) and (\ref{vvi}), we will show that the $V\leftrightarrow S_t$
interchange takes place plane by plane.
The number of vectorials and spinorials originating from
a specific orbifold plane $I$ are interchanged when the
ranks of the associated $Y$-vectors are interchanged
\ba
{\rm rank}\left[\Delta^{(I)},Y_{16}^{(I)}\right]\leftrightarrow{\rm rank}
\left[\Delta^{(I)},Y_{10}^{(I)}\right]~,
\ea
as follows from eqs. (\ref{ssi}) and (\ref{vvi}).
In order to prove the existence of $V\leftrightarrow S_t$ duality
we have to demonstrate the existence of a universal map that preserves
the ranks of the matrices, while exchanging
$Y^I_{16}\leftrightarrow Y^I_{10}$. Since the rank of
the augmented matrix does not change by adding to the
$Y^I_{10}$ a linear combination of the columns of $\Delta^I$
the most general transformations of the  GSO phases that realizes
the above interchange, modulo the rank preserving transformations,
are given by
\ba
\oo{e_k}{b_1} &\rightarrow& \oo{e_k}{b_1+x}+
\sum_{i=3,4,5,6}\lambda_i^1\,\oo{e_k}{e_i}\ ,\ k=1,2\nn\\
\oo{z_m}{b_1} &\rightarrow& \oo{z_m}{b_1+x}+
\sum_{i=3,4,5,6}\lambda_i^1\,\oo{z_m}{e_i}\ ,\ m=1,2\label{tri}
\ea
\ba
\oo{e_k}{b_2} &\rightarrow& \oo{e_k}{b_2+x}+
\sum_{i=1,2,5,6}\lambda_i^2\,\oo{e_k}{e_i}\ ,\ k=3,4\nn\\
\oo{z_m}{b_2} &\rightarrow& \oo{z_m}{b_2+x}+
\sum_{i=1,2,5,6}\lambda_i^2\,\oo{z_m}{e_i}\ ,\ m=1,2\label{trii}
\ea
\ba
\oo{e_k}{b_3} &\rightarrow& \oo{e_k}{b_3+x}+
\sum_{i=1,2,3,4}\lambda_i^3\,\oo{e_k}{e_i}\ ,\ k=5,6\nn\\
\oo{z_m}{b_3} &\rightarrow& \oo{z_m}{b_3+x}+
\sum_{i=1,2,3,4}\lambda_i^3\,\oo{z_m}{e_i}\ ,\ m=1,2\label{triii}
\ea
where $\lambda_i^j=0,1$ are arbitrary coefficients.
The freedom of adding these coefficients amounts to reorganizing the
matter spectrum of the vector representations on each twisted plane.
Adding these coefficients is necessary, as we show below, in order to prove
the existence of a duality map on all three twisted planes. Indeed, the
duality map on the first and second planes can be induced by choosing the
independent phases $(e_{1,2}|b_{1,2})$ and $(z_{1,2}|b_{1,2})$ arbitrarily,
without affecting the $\Delta^I$ matrices. In the third plane, however,
{\it i.e.} for $b_3$ which is composed in terms of $b_1$ and $b_2$,
this freedom a priori is not apparent. Replacing $b_3=b_1+b_2+x$
in eq. (\ref{triii}) we obtain
\ba
\oo{e_k}{b_3} &\rightarrow& \oo{e_k}{b_1}+\oo{e_k}{b_2}+
\sum_{i=1,2,3,4}\lambda_i^3\,\oo{e_k}{e_i}\ ,\ k=5,6\label{triiii}\\
\oo{z_m}{b_3} &\rightarrow& \oo{z_m}{b_1}+ \oo{z_m}{b_2}+
\sum_{i=1,2,3,4}\lambda_i^3\,\oo{z_m}{e_i}\ ,\ m=1,2\label{triiiii}
\ea
The transformation in eq. (\ref{triiii}) is trivially realized
by using the freedom in the phases $\oo{e_5}{b_1}$ and
$\oo{e_6}{b_2}$. Turning to the transformation in eq. (\ref{triiiii}),
there is no remaining freedom in the choice of the GGSO coefficients
$\oo{z_m}{b_1}$ and $\oo{z_m}{b_2}$ since these are used in the
transformations on the first two planes.
Using the duality transformations in eqs. (\ref{tri}-\ref{triii})
we may rewrite eq. (\ref{triiiii})
as
\ba
\oo{z_m}{b_3}&\rightarrow&
 \oo{z_1}{b_3+x}+\left[\oo{z_1}{z_2}+1
+\sum_{i=1,2}\left(\lambda_i^2+\lambda_i^3+1 \right)\oo{z_m}{e_i}\right.\nn\\
&&\ \
+\left.\sum_{i=3,4}\left(\lambda_i^1+\lambda_i^3+1 \right)\oo{z_m}{e_i}
+\sum_{i=5,6}\left(\lambda_i^1+\lambda_i^2+1 \right)\oo{z_m}{e_i}\right]
\label{b3transformation}
\ea
where $m=1,2$ and the identity
\beq
\oo{z_m}{x}= \oo{z_m}{1+S+z_1+z_2+\sum_{1}^{6} e_i}
             =1+\oo{z_1}{z_2}+\sum_1^6\oo{z_m}{e_i}
\eeq
is used to obtain eq. (\ref{b3transformation}). To show the existence
of the duality map on the third twisted plane it is sufficient to
demonstrate that the term in the square brackets in eq.
(\ref{b3transformation}) can be either $0$ or $1$ for appropriate choice
of $\lambda_i^j$ coefficients. This possibility exists provided that
at least one of the $\oo{z_1}{e_i}$ and one of the $\oo{z_2}{e_i}$
is non vanishing. This is indeed the case in the class of models
that we classify here, being the no gauge group enhancement condition
discussed in section \ref{4dgroup}.

\section{Self--dual solutions without enhanced symmetry}\label{sdsolutions}

In this section we discuss the self dual solutions.
The existence of such self dual solutions is evident from
the matrix in figure (\ref{vectorspinormatrix} and the density plot in
figure \ref{densityplot}.
The diagonal elements in the figure and the corresponding matrix
are the self dual solutions, in which
the total number
of $(16\oplus{\overline{16}})$ spinorial representations is equal to
the total number of $(10)$ vectorial representations of $SO(10)$.
This self--duality is obvious when the $SO(10)$ symmetry is enhanced
to $E_6$. Indeed, in this case the 27 contains the $16+10+1$, whereas the
$\overline{27}$ contains the $\overline{16}+10+1$. Hence, in a $E_6$
vacuum with a given number of $27$ and $\overline{27}$ the total number of
$16\oplus\overline{16}$ spinorial representations
is necessarily equal to the total number of
$10$ vectorial representations. However,
in the models that we
classify here the gauge bosons that enhance the $SO(10)\times U(1)$
symmetry to $E_6$ are always projected out by the GGSO projections.
Nevertheless, as illustrated in
\ref{vectorspinormatrix} and \ref{densityplot},
there exist in the space of vacua, models that preserve the
self--duality.

In the case that the symmetry is enhanced to $E_6$, a given
sector $B$, on a given twisted plane,
may give rise to a $16$ or $\overline{16}$
representation of $SO(10)$,
and necessarily an accompanying $10$ vectorial
representation from the sector $B+x$, to supplement the representation to the
$27$ representation of $E_6$. However, once the $E_6$ symmetry is broken
we expect that the given sectors $B$ and $B+x$, give rise to
either a massless spinor or vector, but not to both, and hence that the
equality is removed. Furthermore, as the $E_6$ symmetry is broken we
anticipate that the Abelian $U(1)$ symmetry in
$E_6 \rightarrow SO(10)\times U(1)$ becomes anomalous.
While this expectation is in general correct, there exist models
in the space of vacua in which the total numbers of spinor and vectors
redistribute themselves among the twisted sectors in a way that
maintains the equality of the total number of $(16\oplus\overline{16})$
and $10$ multiplets. In the space of $SO(10)$ vacua classified in our
work, these models are self--dual under the spinor--vector duality.
Furthermore, in some of these self--dual solutions the Abelian $U(1)$
symmetries are anomalous, whereas in others all the Abelian $U(1)$ symmetries
are anomaly free. Below we exhibit two examples of models in this class.

\subsection{A three generation self-dual model}
This model is generated by the basis vectors $\{v_1,\dots,v_{12}\}$ of
(\ref{basis}) and the GGSO coefficients $\cc{v_i}{v_j}=\exp[\oo{v_i}{v_j}]$
, $i,j=1,\dots,12$, where
{$$ (v_i|v_j)\ \ =\ \ \bordermatrix{
&1&S&e_1&e_2&e_3&e_4&e_5&e_6&b_1&b_2&z_1&z_2\cr
 1  & 1& 1& 1& 1& 1& 1& 1& 1& 1& 1& 1& 1\cr
 S  & 1& 1& 1& 1& 1& 1& 1& 1& 1& 1& 1& 1\cr
e_1& 1& 1& 0& 1& 0& 0& 1& 0& 1& 0& 0& 1\cr
 e_2& 1& 1& 1& 0& 1& 0& 1&
0& 0& 0& 0& 0\cr
e_3& 1& 1& 0& 1& 0& 0& 0& 1& 0& 0& 0& 0\cr e_4& 1&
1& 0& 0& 0& 0& 1& 0& 0& 1& 1& 1\cr
 e_5& 1& 1& 1& 1& 0& 1& 0& 1& 1&
1& 1& 1\cr e_6& 1& 1& 0& 0& 1& 0& 1& 0& 1& 0& 1& 0\cr b_1& 1& 0& 1&
0& 0& 0& 1& 1& 1& 0& 1& 1\cr b_2& 1& 0& 0& 0& 0& 1& 1& 0& 0& 1& 1&
1\cr z_1& 1& 1& 0& 0& 0& 1& 1& 1& 1& 1& 1& 1\cr z_2& 1& 1& 1& 0& 0&
1& 1& 0& 1& 1& 0& 1\cr
  }
$$}

The properties that characterize the model are:

$\bullet$ the gauge group is $SO(10)\times{SO(8)}^2\times{U(1)}^3$.

$\bullet$ three $SO(10)$ spinorials
(one from each plane) arising from the points\\
$S+b_1+e_3+e_5$, $S+b_2+e_5$, $S+b_3+e_2+e_4$.

$\bullet$ three $SO(10)$ vectorial representations
(one from each plane) arising from\\
 $S+b_1+e_3+x$, $S+b_2+x$ and
$S+b_3+e_3+x$.

$\bullet$  eight octets charged under the first $SO(8)$ arising from\\
 $S+b_3+e_1+z_2$, $S  +b_3+ e_1  + e_3  + e_4  +x$,
 $S+b_2+e_2+e_5+e_6+z_1+x$,
 $S+b_2+e_2+e_6+x$,
 $S+b_1+e_3+e_4+e_6+z_1+x$,
 $S+e_3+e_4+b_1+z_1+x$,
 $S+b_1+e_3+e_6+x$,
 $S+b_2+e_5+z_1+x$

$\bullet$ eight octets charged under the
second ${SO(8)}$ from\\
$S+b_3+e_3+e_4+z_2+x$,
$S+b_3+e_4+e_5+e_6+x$,
$S+b_3+e_1+e_2+e_3+z_2+x$,
$S+b_2+e_1+e_5+x+z_2$,
$S+b_2+e_1+x$,
$S+b_1+e_3+e_4+e_6+x$,
$S+b_1+e_4+e_5+e_6+x+z_2$,
$S+b_2+x+z_2$.

A number of non-Abelian group singlets is also
present in the model's spectrum. All three Abelian factors are anomaly free.

\subsection{A six generation self-dual model}
This model is generated by the basis vectors
$v_1,\dots,v_{12}$ of (\ref{basis}) and the GGSO coefficients
$\cc{v_i}{v_j}=\exp[\oo{v_i}{v_j}]$
, $i,j=1,\dots,12$, where
{$$ (v_i|v_j)\ \ =\ \ \bordermatrix{
&1&S&e_1&e_2&e_3&e_4&e_5&e_6&b_1&b_2&z_1&z_2\cr
 1  & 1& 1& 1& 1& 1& 1& 1& 1& 1& 1& 1& 1\cr S  & 1& 1& 1& 1& 1& 1& 1&
1& 1& 1& 1& 1\cr e_1& 1& 1& 0& 0& 0& 1& 0& 1& 1& 0& 0& 1\cr e_2& 1&
1& 0& 0& 1& 1& 0& 0& 0& 0& 0& 0\cr e_3& 1& 1& 0& 1& 0& 1& 1& 0& 0&
1& 1& 1\cr e_4& 1& 1& 1& 1& 1& 0& 0& 0& 0& 1& 1& 0\cr e_5& 1& 1& 0&
0& 1& 0& 0& 1& 0& 1& 1& 1\cr e_6& 1& 1& 1& 0& 0& 0& 1& 0& 0& 1& 1&
0\cr b_1& 1& 0& 1& 0& 0& 0& 0& 0& 1& 0& 1& 0\cr b_2& 1& 0& 0& 0& 1&
1& 1& 1& 0& 1& 1& 0\cr z_1& 1& 1& 0& 0& 1& 1& 1& 1& 1& 1& 1& 1\cr
z_2& 1& 1& 1& 0& 1& 0& 1& 0& 0& 0& 0& 1\cr
  }
$$}

The properties that characterize the model are:

$\bullet$ the gauge group is $SO(10)\times{SO(8)}^2\times{U(1)}^3$.

$\bullet$  six $SO(10)$ spinorials (2 from each plane)
arising from the points\\
$S+b_1+e_6, S + b_1+ e_3 + e_4 + e_5, S
+b_2+ e_2 + e_6, S  +b_2+ e_1  + e_5,   S+b_3+e_3+e_4,$\\
$S+ b_3+e_2+e_3+e_4 $.

$\bullet$ six $SO(10)$ vectorials (2 from each plane) arising from\\
 $S+b_1+e_3+e_4+e_6+x$, $S+b_2+e_1+e_6+x$,
$S+b_2+e_2+e_5+x$, $S+b_2+e_5+x$, $S+b_3  + e_1 + e_3 + e_4+x$ , $S
+ b_3+ e_1  + e_2 + e_3 + e_4+x$.

$\bullet$ six octets charged under the first $SO(8)$ arising from\\
$  S  + b_3+e_1    + e_3 +x $, $ S+ b_3+ e_1  + e_2  + e_3  +x$,
$ S + b_2+e_1  + x$, $ S+b_2+e_2+e_5+e_6 + x$,
$ S+b_1+e_3+e_4+e_5+e_6+x+z_1$,
$S+b_1+x+z_1$

$\bullet$ six octets charged under the second ${SO(8)}$ from\\
$S+b_3  + e1  + e4  + x  + z2$,  $S+b_3  + e_1  + e_2  + e_4  + x +
z_2$ ,
$S+b_2+e_1+e_5+e_6+z_2$, $S+b_2+ e_2 + z_2$,
$S+b_1+e_3+3_4+x+z_2$, $S+b_1+e_5+e_6+z_2$.

A number of non-Abelian gauge group singlets is also
present in the model's spectrum. All three Abelian factors are anomaly free.

\section{Conclusions}\label{conclude}

In this paper we continued the classification of fermionic $Z_2\times Z_2$
heterotic string vacua, which was started in ref. \cite{fknr}.
It was restricted in \cite{fknr} to the models dubbed $S^3$ class models,
in which all twisted planes may a priori produce spinorial representations.
Extensions to $S^2V$, $SV^2$ and $V^3$ classes of models by modifying
the set of
boundary condition basis vectors were investigated in ref. \cite{nooij}.
However, as discussed in ref. \cite{fkr} the entire space of $S^3$,
$S^2V$, $SV^2$ and $V^3$ classes of models is generated by using
the single set of basis vectors given in eq. (\ref{basis}) and
modifying the one--loop GGSO projection. This result follows from
theta--functions identities which render the vector basis modification
equivalent to certain choices of GGSO projection coefficients in
the enlarged basis set (\ref{basis}). This equivalence therefore facilitates
the classification of this class of string vacua, as one can work
with a single basis and the classification entails the variation of the
binary GGSO projection coefficients. Counting the number of independent
GGSO phases therefore corresponds to a space of $2^{55}$, or approximately
$10^{16.6}$, independent choices. In ref. \cite{fkr} we resorted to a
random generation of
GGSO phases to scan a space of $\sim10^{10}$ independent
models, with $SO(10)\times U(1)^3\times{hidden}$ gauge group. The hidden
gauge group in that case was not restricted, and the enhancements of
$SO(8)\times SO(8)$ to $SO(16)$ and $E_8$ were allowed.

In the work reported here we
restricted the classification to models in which the hidden gauge group is not
enhanced, and is $SO(10)\times U(1)^3\times SO(8)\times SO(8)$ over the
entire space of models. This restriction reduced the number of
independent phases, and therefore allows the complete
classification of this space of vacua,
and consequently produces exact results.
The classification then reveals a bell shape distribution that peaks
at vanishing net number of chiral families, and $\sim15\%$ of models
with three net chiral families. These results are in accordance with the
statistical results of ref. \cite{fkr}. This outcome
lends credence to recent attempts \cite{statistical}
at using statistical methods to extract phenomenological information on
ensembles of string vacua.

The complete classification revealed a novel duality symmetry over the entire
space of scanned vacua under the exchange of spinorial plus antispinorial
representations of $SO(10)$ with vectorial representations.
This duality symmetry implies that for every model with a given number of
spinors (plus antispinors) and vectors there exist another model in which
they are interchanged, and reflects a symmetry under the discrete exchange
of some GGSO projection coefficients. We exhibited this discrete exchange in
one concrete example and provided a general algebraic proof.

It is important to note that over the space of $Z_2\times Z_2$ heterotic
string vacua that we study here the $S_t\leftrightarrow V$ duality map
operates on the $Z_2\times Z_2$ twisted planes, plane by plane.
As the $Z_2\times Z_2$ twisted planes preserve $N=2$ space--time
supersymmetry this fact implies that the $S_t\leftrightarrow V$
duality already exists at the $N=2$ level. It is of interest therefore
to explore whether the $S_t\leftrightarrow V$ duality also exists
in other classes of string compactification do not contain $N=2$
preserving sectors.

The existence of the duality symmetry over the entire space of vacua is
of fundamental significance. It reflects the existence of a common structure
that underlies the entire space of models. Just as in the case of
ten dimensional string theories and eleven dimensional supergravity,
the existence of nontrivial duality relations suggests
the existence of an underlying theoretical formalism,
traditionally dubbed M--theory, the spinor--vector duality indicates
a common structure that underlies the entire space of
fermionic $Z_2\times Z_2$ vacua. Thus, the view of this
space of string models as consisting of disconnected vacua
is premature, and they may in fact be connected by a yet unknown physical
mechanism.

It is of further interest to develop a geometrical
correspondence of the spinor--vector duality that we uncovered
in the free fermionic, or orbifold, limit. In this respect
the spinor--vector duality may be viewed as a generalization of
the mirror symmetry \cite{mirror},
which exchanges spinors with antispinors.
In the geometrical picture,
just as mirror symmetry indicated the existence
of topology changing transitions between Calabi--Yau manifolds
with a mirror Euler characteristic, but equal in absolute
value \cite{agm}, the spinor--vector duality might indicate the existence
of topology changing transitions between heterotic string vacua with different
Euler character. The geometrical picture in this case, however,
might prove more intricate to explore as one must also take account of the
vector bundle that accounts for the heterotic string gauge
degrees of freedom. Nevertheless, the feasibility of such
transitions, afforded by the observation of the spinor--vector
duality over the entire space of fermionic $Z_2\times Z_2$ vacua,
suggests that the models in this space are connected by a yet
unknown mechanism rather than disconnected.

\section{Acknowledgments}

We would like to CERN  theory division
for hospitality.
AEF would like to thanks the Oxford theory department for hospitality
and is supported in part by PPARC under contract PP/D000416/1.
CK is supported in part by the EU under contracts
MTRN--CT--2004--005104, MTRN--CT--2004--512194 and
ANR (CNRS--USAR) contract No 05--BLAN--0079--01 (01/12/05)
JR is supported by the program ``PYTHAGORAS''  (no. 1705 project 23)
of the Operational Program for Education and Initial Vocational
Training of the Hellenic Ministry of Education under the
3rd Community Support Framework and the European Social Fund;
and by the EU under contract MRTN--CT--2004--503369.



\bigskip
\medskip

\bibliographystyle{unsrt}

\end{document}